\documentclass[fleqn,usenatbib]{mnras}

\usepackage[T1]{fontenc}
\usepackage{ae,aecompl}

\usepackage{graphicx}	
\usepackage{amsmath}	
\usepackage{amssymb}	

\usepackage{newtxtext,newtxmath}

\title[When will T CrB next erupt as a nova?]{When will T Coronae Borealis next erupt as a nova? Constraints from recurrence, orbital phase, and accretion-state evolution}

\author[Pei et al.]{
Songpeng Pei,$^{1}$\thanks{E-mail: songpengpei@outlook.com}
Xiaowan Zhang,$^{1}$
Renzhi Su,$^{2}$
Ziwei Ou,$^{3}$
Yongzhi Cai,$^{4, 5}$
Qiang Li,$^{6, 7}$
\newauthor
Yu Liu,$^{8}$
Xiaoqin Ren,$^{6, 7}$
and Taozhi Yang$^{9}$
\\
$^{1}$School of Physics and Electrical Engineering, Liupanshui Normal University, Liupanshui, Guizhou, 553004, China\\
$^{2}$Shanghai Astronomical Observatory, Chinese Academy of Sciences, 80 Nandan Road, Shanghai 200030, China\\
$^{3}$Tsung-Dao Lee Institute, Shanghai Jiao Tong University, Shanghai 201210, China\\
$^{4}$International Centre of Supernovae (ICESUN), Yunnan Key Laboratory of Supernova Research, Yunnan Observatories, Chinese Academy of Sciences (CAS),\\ Kunming, 650216, China\\
$^{5}$INAF - Osservatorio Astronomico di Padova, Vicolo dell'Osservatorio 5, 35122 Padova, Italy\\
$^{6}$School of Physics and Electronic Science, Qiannan Normal University for Nationalities, Duyun 558000, China\\
$^{7}$Qiannan Key Laboratory of Radio Astronomy, Guizhou Province, Duyun 558000, China\\
$^{8}$State Key Laboratory of Public Big Data, Guizhou University, Guiyang 550025, China\\
$^{9}$Ministry of Education Key Laboratory for Nonequilibrium Synthesis and Modulation of Condensed Matter, School of Physics, Xi'an Jiaotong University,\\ 710049 Xi'an, China\\
}

\date{Accepted 00.00.2026. Received 00.00.2026; in original form 00.00.2026}

\pubyear{2015}

\begin{document}
\label{firstpage}
\pagerange{\pageref{firstpage}--\pageref{lastpage}}
\maketitle

\begin{abstract}
T Coronae Borealis (T CrB) is the nearest symbiotic recurrent nova and is now intensively monitored for its next eruption. We combine three constraints on the eruption time: historical recurrence, orbital phase, and recent accretion-state evolution. Conditioning on no eruption by 2026 July 11, the three effective historical intervals give illustrative survival-conditioned probabilities of 30.2\% for the rest of 2026 and 56.9\% within the following year; these are empirical indicators, not physical prediction probabilities. The four adopted historical eruption phases do not select a unique ignition phase, but form two loose pairs near $\phi\simeq0.44$ and $\phi\simeq0.62$, used here only as monitoring windows. The 1946 pre-eruption dip is difficult to explain by either a pure accretion-rate decline or standard dust extinction, and may have involved both accretion restructuring and source-dependent obscuration. If the renewed 2026 decline is the true pre-1946 analogue, an eruption around 2026 December remains plausible. Conversely, if post-2024 brightness remains below the 2014--2023 high state, an accretion-deficit estimate gives an earliest lower limit near 2029 May. The data therefore support conditional monitoring windows, not a unique date. These are conditional diagnostic scenarios rather than competing point predictions; the eventual eruption epoch will test their underlying assumptions.

\end{abstract}



\section{Introduction}
\label{sec:intro}

T CrB (HR~5958; HD~143454) is the closest known recurrent nova (RN) and a prototype of the class. Its 1866 and 1946 eruptions both reached approximately $V\simeq2$~mag \citep{1949ApJ...109...81S}, and possible earlier events have been proposed for AD~1217 and 1787 \citep{2023JHA....54..436S,2023ATel16107....1S}. About 80~yr after the 1946 eruption, the system is widely regarded as approaching its next nova outburst \citep{2016NewA...47....7M,2023ApJ...953L...7I,2023AstL...49..501M,2023A&A...680L..18Z,2025MNRAS.541L..14M,2026A&A...706A..94P,2026A&A...707A.102L}. Published estimates have placed the eruption broadly in 2023.0--2026.8 \citep{2020ApJ...902L..14L,2023ATel16107....1S,2023MNRAS.524.3146S}, but its exact timing remains uncertain \citep{2025MNRAS.541L..14M, 2026A&A...707A.102L}.

The binary contains a Roche-lobe-filling M4~III red giant and a massive white dwarf (WD) \citep{1998MNRAS.296...77B,1999A&AS..137..473M}. Recent work gives $M_{\rm RG}=0.69^{+0.02}_{-0.01}\,M_{\odot}$ and $R_{\rm RG}=65\pm6\,R_{\odot}$ \citep{2025ApJ...983...76H}; published WD masses span $1.32$--$1.38\,M_{\odot}$ \citep{2004A&A...415..609S,2018ApJ...860..110S,2019ApJS..242...18H,2025ApJ...983...76H,2025A&A...701A.176M}. The orbital period is tightly constrained to $P_{\rm orb}\simeq227.55$--$227.58$~d \citep{2000AJ....119.1375F,2023MNRAS.524.3146S,2025A&A...694A..85P,2025ApJ...983...76H,2025A&A...701A.176M,2025ApJ...991..111S}, with ellipsoidal optical variability at half this period \citep{1975JBAA...85..217B,2023MNRAS.524.3146S}. The {\it Gaia} eDR3 distance is $896\pm22$~pc\footnote{http://dc.g-vo.org/tableinfo/gedr3dist.main} \citep{2021AJ....161..147B}, consistent with the $\sim914$~pc estimate of \citet{2022MNRAS.517.6150S}. T CrB also shows long active phases that may be dwarf-nova-like \citep{2023ApJ...953L...7I}.

The recent activity has been compared with the pre-1946 behaviour. T CrB entered a high or super-active state around 2014--2015, similar to the enhanced state before the 1946 nova \citep{2016NewA...47....7M,2016MNRAS.462.2695I,2018A&A...619A..61L}, and this was interpreted as enhanced accretion and a possible precursor \citep{2016NewA...47....7M,2020ApJ...902L..14L,2023MNRAS.524.3146S,2023ApJ...953L...7I,2023A&A...680L..18Z}. We call this interval the ``high state''. The following 2023--2024 fading episode was likewise proposed as a pre-eruption dip because a similar decline occurred before 1946 \citep{2023MNRAS.524.3146S,2023ATel16107....1S,2023BAAVC.196....8T,2023ATel16114....1K}.

The predictive value of this analogy is uncertain. \citet{2025MNRAS.541L..14M} showed that the 2023--2024 dip matches the pre-1946 event better in $B$ than in $V$ and should not be used as a reliable eruption clock. Multiwavelength studies also favour an accretion-flow and boundary-layer change over dust obscuration \citep{2026A&A...706A..94P,2026A&A...707A.102L}. Thus the relevant question is not which single date is correct, but what constraints can be obtained from recurrence, orbital phase, and accretion evolution.

The historical recurrence record is also useful but limited. \citet{2025ApJ...991..111S} argued that the recurrence timescale may have shortened by about $-0.45$~yr per eruption, whereas the interval from 1946 February to 2026 July 11 already exceeds 80.41 yr, longer than the 79.75~yr 1866--1946 interval. The recurrence pattern of Galactic recurrent novae shows scatter of at least 10~yr around the mean $\sim$80~yr cycle of T CrB \citep{2026A&A...707A.102L}; recurrence can therefore define a broad window, not a deterministic date.

In this Letter, we combine three constraints without treating any one observable as a clock: (1) historical recurrence intervals, used for survival-conditioned, small-sample probabilities; (2) historical eruption phases, used only as empirical monitoring windows; and (3) recent optical evolution, used with published accretion-state interpretations to assess whether the shorter and fainter 2014--2023 high state can explain the absence of an eruption by 2026. We formulate these constraints as falsifiable scenarios: after the eruption, its measured epoch will remain useful for assessing the recurrence model, the proposed dip analogue, the phase-window construction, and the accretion-deficit assumptions.

\section{Observations and Data}
\label{sec:observation}

We use Johnson $B$- and $V$-band photometry of T CrB from the American Association of Variable Star Observers (AAVSO) International Database. The data span MJD 52713--61232 (2003 March 15--2026 July 11), ending with the most recent AAVSO measurements. Following \citet{2023MNRAS.524.3146S}, we binned both light curves into 0.01~yr intervals; the results are shown in Figure~\ref{fig:lc}.

This interval covers the pre-2014 quiescent baseline, the 2014--2023 high state, the 2023--2024 fading episode, the post-dip recovery after 2024 May, and the renewed decline visible in 2026. We use these light curves mainly for timing and state classification. Their relative brightness levels provide the basis for the accretion-deficit estimate in Section~\ref{subsec:accretion_deficit_timescale} and for judging whether the current decline could be a closer analogue of the pre-1946 dip.

\section{Analysis}
\label{sec:analysis}

\subsection{Conditional recurrence probability}
\label{sec:chance}

The adopted eruption epochs \citep{2024RNAAS...8..272S} give three effective intervals of 81.44, 78.39, and 79.75~yr (the 1217--1787 separation represents seven cycles). Using Julian years, their mean and sample standard deviation are $\bar{T}=79.86$~yr and $s_T=1.53$~yr; by 2026 July 11 (JD~2461232.5), the post-1946 interval was 80.415~yr.

We describe these three values by a Gaussian with $\mu=\bar{T}$ and $\sigma=s_T$. If $F(T)$ is its cumulative distribution, conditioning on no eruption by $t_0$ gives
\begin{equation}
P(t_0<T<t_1\,|\,T>t_0)=
\frac{F(t_1)-F(t_0)}{1-F(t_0)} .
\label{eq:conditional_probability}
\end{equation}
For $t_0=$ 2026 July 11, the probabilities are 30.2\% through 2026 and 56.9\% over the following 12 months. The instantaneous hazard
\begin{equation}
h(t)=\frac{f(t)}{1-F(t)}
\label{eq:hazard}
\end{equation}
is $0.682~{\rm yr}^{-1}$ at $t_0$ and $0.831~{\rm yr}^{-1}$ at the end of 2026 if no eruption occurs. Leave-one-out ranges are 20--64\%, 40--91\%, and $0.44$--$1.94~{\rm yr}^{-1}$, respectively; these are sensitivity ranges, not confidence intervals, and exclude uncertainties in the AD~1217 and 1787 dates. They confirm that recurrence is an empirical indicator, not a physical prediction.

\citet{2024RNAAS...8..272S} also noted near-integer orbital-cycle separations. The 2026 June 25 candidate has passed; the next is 2027 February 8. With no physical triggering mechanism, these dates are monitoring markers only.

\subsection{Empirical orbital-phase monitoring windows}
\label{subsec:phase_windows}

We fold the same four eruptions on a fixed spectroscopic ephemeris to define monitoring windows, without assuming phase-triggered ignition.

We compute the orbital phase using the ephemeris of \citet{2000AJ....119.1375F},
\begin{equation}
P_{\rm orb}=227.5687~{\rm d},\qquad
T_0={\rm HJD}~2447918.62 ,
\end{equation}
with
\begin{equation}
\phi = {\rm frac}\left[\frac{t-T_0}{P_{\rm orb}}\right],
\label{eq:phase_definition}
\end{equation}
where ${\rm frac}$ denotes the fractional part of the cycle.
For the four historical eruption epochs, this gives
\begin{equation}
\begin{aligned}
\phi_{1217}&=0.6177,\quad \phi_{1787}=0.6293,\\
\phi_{1866}&=0.4462,\quad \phi_{1946}=0.4384 .
\end{aligned}
\end{equation}
These values do not define a single preferred eruption phase.
Rather, they separate into two loose empirical pairs: one near $\phi\simeq0.44$, corresponding to the 1866/1946 pair, and one near $\phi\simeq0.62$, corresponding to the 1217/1787 pair.
The corresponding pair-averaged phases are
\begin{equation}
\bar{\phi}_{0.44}=0.4423,\qquad
\bar{\phi}_{0.62}=0.6235 .
\end{equation}

For a future epoch with phase $\phi$, the possible observing dates are
\begin{equation}
t(\phi,N)=T_0+(N+\phi)P_{\rm orb},
\label{eq:phase_epoch}
\end{equation}
where $N$ is an integer chosen so that $t$ falls after the reference date.
Applying Equation~(\ref{eq:phase_epoch}) to the two empirical phase pairs gives the remaining monitoring windows after 2026 July 11 through the end of 2030. The first remaining window is 2026 August 5--8, and the next complete pair-spanning interval is 2027 February 8--March 23.

These dates should not be interpreted as deterministic predictions.
The historical sample contains only four events, the AD~1217 and 1787 dates are uncertain, and previous phase-distribution tests have not established statistically robust orbital-phase locking \citep{2026arXiv260520991P}.
The dates in Table~\ref{tab:phase_windows} are therefore best regarded as empirical observing windows: useful for planning dense monitoring, but not evidence that the eruption must occur at one of these phases. The widths are pair separations, not confidence intervals; uncertain early dates and long-term period evolution dominate the formal ephemeris error.

\begin{table*}
\centering
\caption{Empirical orbital-phase monitoring windows for T CrB after 2026 July 11 through 2030. 
The dates are computed by folding the historical eruption epochs on the fixed spectroscopic ephemeris of \citet{2000AJ....119.1375F} and using the two loose phase pairs found in the historical record: $\phi\simeq0.44$ for the 1866/1946 pair and $\phi\simeq0.62$ for the AD~1217/1787 pair.
The windows are intended for observational planning only and should not be interpreted as physically required eruption dates.}
\label{tab:phase_windows}
\begin{tabular}{lcccc}
\hline
Window & $\phi\simeq0.44$ pair window & $\bar{\phi}_{0.44}=0.4423$ date
& $\phi\simeq0.62$ pair window & $\bar{\phi}_{0.62}=0.6235$ date \\
\hline
1 & -- & -- & 2026 Aug 5--8 & 2026 Aug 6--7 \\
2 & 2027 Feb 8--10 & 2027 Feb 9 & 2027 Mar 21--23 & 2027 Mar 22 \\
3 & 2027 Sep 24--25 & 2027 Sep 24--25 & 2027 Nov 3--6 & 2027 Nov 5 \\
4 & 2028 May 8--10 & 2028 May 9 & 2028 Jun 18--21 & 2028 Jun 19--20 \\
5 & 2028 Dec 22--23 & 2028 Dec 23 & 2029 Jan 31--Feb 3 & 2029 Feb 2 \\
6 & 2029 Aug 6--8 & 2029 Aug 7 & 2029 Sep 16--19 & 2029 Sep 17--18 \\
7 & 2030 Mar 22--24 & 2030 Mar 23 & 2030 May 2--4 & 2030 May 3 \\
8 & 2030 Nov 4--6 & 2030 Nov 5 & 2030 Dec 15--18 & 2030 Dec 16--17 \\
\hline
\end{tabular}
\end{table*}

\subsection{The 1946 pre-eruption dip and the 2023--2024 fading episode}
\label{subsec:dip}

The physical origin of the pre-1946 dip is uncertain. \citet{2023MNRAS.524.3146S} suggested circumstellar dust from a previous disk ejection, but a standard Galactic extinction law gives $A_B>A_V$ \citep[e.g.][]{1989ApJ...345..245C,1999PASP..111...63F}, whereas the historical record has been interpreted as showing a deeper decline in $V$ than in $B$. Dust is also disfavoured for the recent faint state by the lack of strong infrared excess and by ultraviolet dust-sensitive diagnostics \citep{2026A&A...707A.102L}.

Other explanations are incomplete. An increased inner disk radius before thermonuclear runaway \citep{2025ApJ...989...78S} can reduce blue/UV inner-disk light, and a lower accretion rate with a boundary-layer transition explains the modern multiwavelength behaviour \citep{2026A&A...706A..94P}. Neither mechanism alone explains a $V$ level below the expected red-giant contribution, or a decline deeper in $V$ than in $B$, if the historical depth is physical.

For a broad-band flux decomposed into red-giant and accretion components \citep[e.g.][]{2023ApJ...953L...7I,2026A&A...706A..94P}, a pure hot-component decline has a lower limit. In the $V$ band,
\begin{equation}
F_V=F_{V,\rm RG}+F_{V,\rm acc},
\end{equation}
and therefore
\begin{equation}
F_V\rightarrow F_{V,\rm RG}
\quad {\rm as}\quad
F_{V,\rm acc}\rightarrow0 .
\end{equation}
Thus, lowering the accretion luminosity or truncating the hot inner disk can make the system approach, but not fall below, the red-giant level unless some cool optical light is also attenuated.

We therefore regard the 1946 dip, if robust, as a composite event: the accretion component faded or was restructured while an asymmetric, optically thick, or grey circumstellar structure partly obscured the red giant or the $V$-dominant continuum. A schematic partial-covering form is
\begin{equation}
F_{V,\rm obs}
=
(1-C_{\rm RG})F_{V,\rm RG}
+
(1-C_{\rm acc,V})F_{V,\rm acc,V},
\end{equation}
and
\begin{equation}
F_{B,\rm obs}
=
(1-C_{\rm RG})F_{B,\rm RG}
+
(1-C_{\rm acc,B})F_{B,\rm acc,B}
+
F_{B,\rm sc/em}.
\end{equation}
Here $C_{\rm RG}$ and $C_{\rm acc}$ are effective covering factors, while $F_{B,\rm sc/em}$ represents scattered blue light, Balmer-continuum emission, line contamination, or residual accretion light. A deeper $V$ decline then requires stronger covering of the red giant/cool continuum or an additional blue component compensating the $B$ attenuation.

Possible realizations include an asymmetric pre-eruption outflow, disk/envelope structure, puffed-up outer rim, stream-overflow material, or large-grain/grey dust in a non-spherical geometry. Such material could obscure the red giant or cooler optical continuum more efficiently than the compact blue-emitting region, while fading or truncation of the hot component explains the $B$ decline. The geometry and formation mechanism remain uncertain.

This interpretation is conservative because the 1946 dip is reconstructed from historical visual and photographic data, not modern simultaneous CCD photometry. Visual-to-$V$ transformations, color evolution, comparison-star sequences, emission lines, and observer response may affect the inferred depth. Thus, if the extreme $V$ minimum is physical, both accretion restructuring and source-dependent obscuration are required; if not, part of the anomaly may be photometric systematics. In either case, the 2023--2024 fading episode is not a strict one-to-one analogue of the 1946 dip. Here, ``accretion restructuring'' denotes changes that reduce or redistribute hot accretion luminosity, including a lower accretion rate, a boundary-layer transition, or an increased inner disk radius.

\subsection{A renewed decline in 2026}
\label{subsec:decline}

After the 2023--2024 fading episode, T CrB brightened again around 2024 May, but by 2026 July 11 the AAVSO light curve showed a renewed decline (Figure~\ref{fig:lc}). The $B$-band brightness had approached the pre-high-state quiescent level, whereas the $V$-band brightness had declined slightly below both its pre-high-state quiescent level and the level reached during the 2023--2024 fading episode.

If this fading continues, it may represent the final decay of the post-dip recovery and could become a closer photometric analogue of the pre-1946 dip than the 2023--2024 episode. Because the interval between the end of the pre-1946 high state, approximately the start of the historical dip, and the 1946 eruption was about six months, a continuing decline would leave open a near-term eruption around 2026 December. Because neither the onset of the present decline nor the historical analogy is sharply defined, the timing uncertainty is at least several months. This is only a monitoring hypothesis: the light curve shows continuing activity, not a deterministic countdown.

\subsection{An accretion-deficit timescale estimate}
\label{subsec:accretion_deficit_timescale}

The delayed eruption of T CrB can also be expressed in a mass-budget form.
A nova eruption occurs when the accreted hydrogen-rich envelope on the WD reaches the ignition mass, $M_{\rm ign}$ \citep[e.g.][]{1982ApJ...257..752F,1982ApJ...257..767F,2005ApJ...623..398Y}.
If the present cycle has not yet supplied the required envelope mass, the remaining waiting time may be written schematically as
\begin{equation}
\Delta t_{\rm wait}
\simeq
\frac{\Delta M_{\rm def}}{\dot{M}_{\rm future}},
\label{eq:wait_time_general}
\end{equation}
where $\Delta M_{\rm def}$ is the remaining mass deficit relative to ignition and $\dot{M}_{\rm future}$ is the future accretion rate onto the WD.
Although $\Delta M_{\rm def}$ is not directly measured, it can be estimated at order-of-magnitude level by comparing the recent high state with the high state preceding the 1946 eruption.

\citet{2025A&A...701A.176M} argued that the 2014--2023 SAP was caused by an inside-out collapse of the accretion disk, during which the mean accretion rate onto the WD was about 28 times larger than in quiescence.
We therefore write
\begin{equation}
\dot{M}_{\rm h}=28\dot{M}_{\rm q},
\label{eq:Mdot_high}
\end{equation}
where $\dot{M}_{\rm q}$ is the quiescent accretion rate and $\dot{M}_{\rm h}$ is the characteristic high-state accretion rate.
Using the onset and termination times measured from the optical light curve \citep{2026arXiv260520991P}, the duration of the high state preceding the 1946 eruption was
\begin{equation}
D_{1946}=31540-27875=3665~{\rm d},
\end{equation}
whereas the duration of the recent 2014--2023 high state was
\begin{equation}
D_{\rm cur}=60126-56795=3331~{\rm d}.
\end{equation}

The optical luminosity of the hot component can be used as a first-order tracer of the accretion rate.
Following \citet{2023A&A...680L..18Z}, we write
\begin{equation}
L = \frac{1}{2}\frac{G M_{\rm WD}\dot{M}_{\rm a}}{R_{\rm WD}}\cos i ,
\label{eq:acc_lum}
\end{equation}
where $G$ is the gravitational constant, $M_{\rm WD}$ and $R_{\rm WD}$ are the WD mass and radius, $\dot{M}_{\rm a}$ is the accretion rate, and $i$ is the inclination.
For fixed $M_{\rm WD}$, $R_{\rm WD}$, and $i$, this gives $\dot{M}_{\rm a}\propto L$ and $M_{\rm acc}\propto\int L(t)\,dt$.
The lower radiated brightness of the 2014--2023 SAP relative to the pre-1946 SAP therefore suggests a smaller integrated mass flow through the disk and onto the WD, provided that the bolometric correction and radiative efficiency were comparable in the two cycles.

The estimate remains highly approximate.
The observed optical luminosity is not necessarily the bolometric accretion luminosity, and changes in boundary-layer optical depth, disk geometry, inclination-dependent projection, extinction, and the unobserved extreme ultraviolet (EUV)/X-ray contribution can all affect the conversion from optical brightness to $\dot{M}_{\rm a}$.
Nevertheless, as an illustrative estimate, \citet{2025A&A...701A.176M} found that the overall radiated brightness of the high state preceding the 1946 eruption was about 40\% larger than that of the recent one.
If the pre-1946 high state is approximated as having a mean accretion rate larger by a factor $f_{\rm b}=1.4$, then the missing high-state-equivalent exposure is
\begin{equation}
\Delta t_{\rm h,eq}=f_{\rm b}D_{1946}-D_{\rm cur}.
\label{eq:delta_t_high_equiv}
\end{equation}
Substitution gives
\begin{equation}
\Delta t_{\rm h,eq}
=
1.4\times3665-3331
=
1800~{\rm d}
=
4.93~{\rm yr}.
\label{eq:delta_t_high_equiv_value}
\end{equation}

If the future accretion rate is
\begin{equation}
\dot{M}_{\rm future}=r\dot{M}_{\rm q},
\end{equation}
the time needed to compensate the deficit is
\begin{equation}
\Delta t_{\rm wait}
=
\frac{\dot{M}_{\rm h}\Delta t_{\rm h,eq}}
     {\dot{M}_{\rm future}}
=
\frac{28}{r}\Delta t_{\rm h,eq}.
\label{eq:future_wait_time}
\end{equation}
Table~\ref{tab:accretion_deficit} lists three illustrative cases.

\begin{table}
\begin{minipage}{80mm}
\caption{Illustrative waiting times required to compensate the estimated high-state accretion deficit.
The calculation assumes $\dot{M}_{\rm h}=28\dot{M}_{\rm q}$ and $\Delta t_{\rm h,eq}=1800~{\rm d}=4.93~{\rm yr}$.
The values are order-of-magnitude timescales, not predictions of the eruption date.}
\label{tab:accretion_deficit}
\begin{tabular}{lccc}
\hline
Assumed future accretion rate & $r$ & $\Delta t_{\rm wait}$ & Interpretation \\
\hline
$\dot{M}_{\rm future}=\dot{M}_{\rm h}$ 
& 28 
& 4.93
& continued high state \\
$\dot{M}_{\rm future}=(\dot{M}_{\rm q}+\dot{M}_{\rm h})/2$ 
& 14.5 
& 9.52
& intermediate state \\
$\dot{M}_{\rm future}=\dot{M}_{\rm q}$ 
& 1 
& 138.0 
& quiescence \\
\hline
\end{tabular}
\end{minipage} 
\end{table}

The WD-mass evolution inferred by \citet{2025ApJ...991..111S} provides an additional qualitative consideration.
They estimated an ejecta mass of
\begin{equation}
M_{\rm ejecta}=0.00074\pm0.00009\,M_{\odot},
\end{equation}
much larger than the accreted mass inferred for the 1866--1946 cycle.
This suggests net WD mass loss, which would increase the ignition mass in a simple like-for-like comparison.
Following the standard thin-shell hydrostatic scaling for nova ignition \citep[e.g.][]{1982ApJ...257..752F,1982ApJ...257..767F,2004ApJ...600..390T,2005ApJ...623..398Y},
\begin{equation}
P_{\rm base}\simeq
\frac{G M_{\rm WD}M_{\rm env}}{4\pi R_{\rm WD}^{4}},
\end{equation}
so that for fixed ignition pressure
\begin{equation}
M_{\rm ign}
\simeq
\frac{4\pi R_{\rm WD}^{4}P_{\rm ign}}
     {G M_{\rm WD}}
\propto
\frac{R_{\rm WD}^{4}}{M_{\rm WD}} .
\end{equation}
Detailed nova calculations show that the ignition mass also depends on the WD core temperature, accretion rate, envelope composition, and mixing history \citep[e.g.][]{1982ApJ...257..752F,1982ApJ...257..767F,2004ApJ...600..390T,2005ApJ...623..398Y,2009ApJ...692..324S}.
For a WD mass near $1.35\,M_{\odot}$, using a standard WD mass--radius relation \citep[e.g.][]{1972ApJ...175..417N}, the mass decrease implied by the 1946 cycle changes $M_{\rm ign}$ by only of order 1--2\% in this estimate.

Therefore, the WD-mass decrease alone is unlikely to delay the eruption by many years.
Applied to the three illustrative accretion-deficit timescales in Table~\ref{tab:accretion_deficit}, a representative 1.8\% ignition-mass correction gives approximately 5.02~yr, 9.69~yr, and 57028~d = 140.5~yr.
The first value, 5.58~yr, is the relevant lower-limit timescale if the future accretion rate does not exceed the 2014--2023 high-state value.
The larger uncertainty remains the amount of mass actually delivered during the 2014--2023 high state and after 2024.

If the post-2024 accretion proceeds at the same mean rate as the 2014--2023 high state, the estimated compensation time is 1800~d, or 4.93~yr.
Taking the renewed inner-disk accretion from 2024 May as the starting point \citep{2025A&A...701A.176M}, this would point to a timescale around spring 2029.
At lower accretion rates, the waiting time is substantially longer.
Thus, a near-term eruption requires either that the remaining ignition mass is small or that the post-2024 accretion rate remains well above the ordinary quiescent value.

\begin{figure*}
\centering
\includegraphics[width=\textwidth]{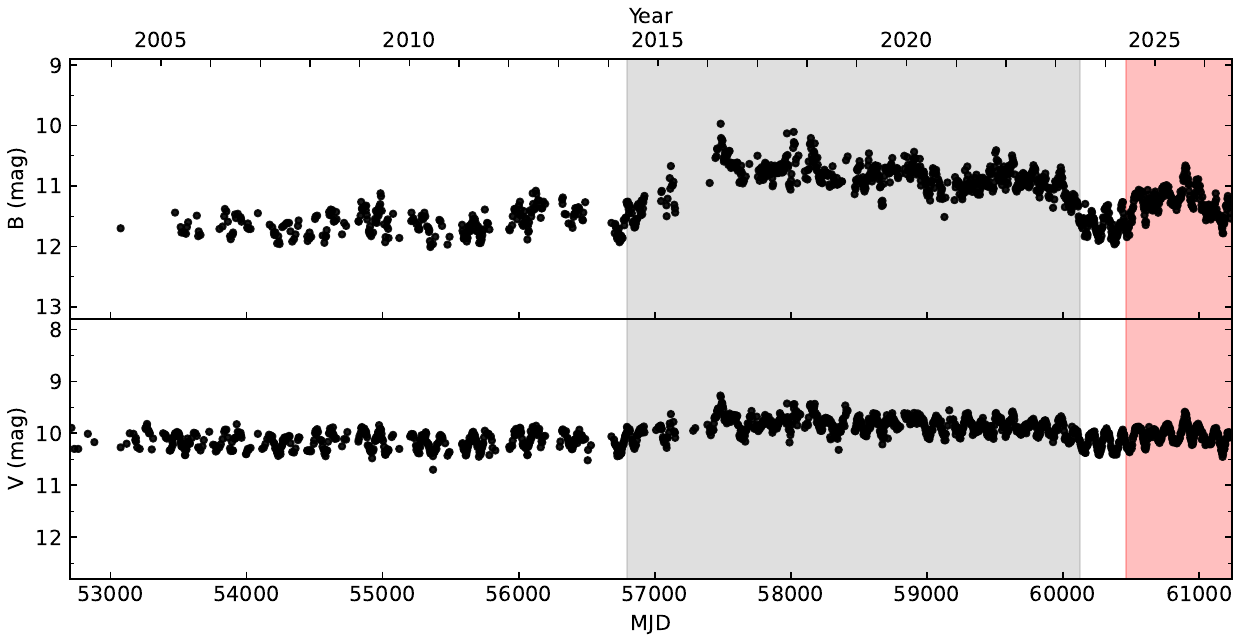}
\caption{AAVSO $B$- and $V$-band light curves of T CrB from MJD~52713 (2003 March 15) to MJD 61232 (2026 July 11), binned in 0.01~yr intervals.
The upper panel shows the $B$ band and the lower panel shows the $V$ band; the magnitude axis is inverted so that brighter states appear higher.
The gray shaded region marks the 2014--2023 high state, and the red shaded region marks the post-dip phase after 2024 May discussed in the text.}
\label{fig:lc}
\end{figure*}

As seen in Figure~\ref{fig:lc}, the mean brightness during the 2014--2023 high state is higher than that during the post-dip phase after 2024 May, implying that the mean accretion rate during the 2014--2023 high state was higher than that during the post-dip phase.
At the same time, the post-dip phase remains brighter on average than the quiescent state before the 2014--2023 high state, implying that the post-dip accretion rate is still above the ordinary quiescent value.
If the mean brightness from 2024 May to the end of the light curve on MJD~61232, together with the future mean brightness, remains lower than during the 2014--2023 high state, then the continued-high-state case provides an earliest-time lower limit rather than a central prediction.
Including the small WD-mass correction discussed above changes the fiducial 4.93~yr estimate to $5.02\pm2.08$~yr; therefore, the corresponding conditional epoch is around 2029 May, with an approximate systematic range spanning 2027--2031. It should not be interpreted as a sharp lower bound.
In this conditional case, the next nova eruption of T CrB would be more likely to occur near or after the fiducial 2029 May epoch, subject to the large systematic uncertainty above.

\section{Discussion}
\label{sec:discussion}

The results above argue against using any single observable as a precise eruption clock for T CrB. The recurrence record places the system in the correct broad eruption interval, but the historical sample is too small for a narrow probability distribution. The orbital phases identify useful monitoring windows, but not a unique ignition phase. The 2023--2024 fading episode demonstrates a major accretion-state transition, yet its mismatch with the 1946 dip, especially in $V$, weakens its use as a strict countdown marker.

The accretion-state evolution provides a natural explanation for why an eruption by 2026 is not required. In symbiotic recurrent novae, much of the ignition mass may be accumulated during high-accretion states rather than during long quiescent intervals \citep{2020ApJ...902L..14L}. For T CrB, the recent high state appears shorter and fainter than the one preceding 1946 \citep{2025A&A...701A.176M,2026arXiv260520991P}. If optical/ultraviolet (UV) luminosity traces the mass delivered to the WD, the 2014--2023 high state may have left an accretion deficit, making the absence of a nova by 2026 physically understandable.

This conclusion does not exclude a 2026 eruption. The deficit estimate is deliberately schematic: the remaining ignition mass may be smaller than implied by the high-state comparison, and the bolometric correction, EUV luminosity, boundary-layer state, disk radiative efficiency, and fraction of disk mass reaching the WD may differ between cycles. Renewed high accretion after 2024 could also reduce the remaining waiting time. Thus, a 2026 eruption remains possible, but the present data do not require it.

The renewed decline seen in 2026 is particularly important for monitoring. If it continues and becomes deeper than the 2023--2024 dip, it may be a closer analogue of the pre-1946 fading episode; an eruption about six months later, around 2026 December, with an uncertainty of at least several months, would support the historical-template interpretation. Conversely, continued post-SAP fluctuations without eruption would favour the view that the high state and dip are imperfect tracers of the ignition condition.

The different timing estimates are conditional diagnostics, not mutually exclusive point predictions. The recurrence-time distribution supplies a statistical prior; the orbital-phase windows are scheduling markers without a triggering mechanism; the dip analogy tests whether a specific light-curve morphology is predictive; and the accretion-deficit calculation tests whether optical luminosity traces the mass still required for ignition. Therefore, verification of one date does not automatically make every other calculation ``wrong'', but it changes the weight assigned to their assumptions.

An eruption during the present decline, particularly around 2026 December, would support the dip-analogue interpretation and imply that the accretion deficit was overestimated, that the post-2024 accretion rate was higher than inferred optically, or that little ignition mass remained. An eruption before the fiducial 2029 epoch but outside the six-month window would weaken the dip as a clock while still requiring a smaller deficit or more efficient accretion than the fiducial estimate. An eruption near or after 2029, provided that the post-2024 mean brightness remains below the 2014--2023 high-state level, would lend support to the deficit interpretation. Coincidence with an orbital-phase window alone would remain weak evidence, because one additional eruption cannot establish phase locking.

Thus, the paper retains diagnostic value after the eruption: the measured eruption epoch will update the recurrence distribution and hazard, add one phase to the historical sample, test the proposed dip delay, and constrain the mapping between radiated high-state luminosity and accumulated ignition mass. We recommend treating the Table~\ref{tab:phase_windows} dates as practical observing windows rather than predictions. Dense optical, ultraviolet, X-ray, and spectroscopic monitoring through these windows and during the ongoing decline remains the most defensible strategy.

\section{Summary and Conclusions}
\label{sec:conclusions}

We have examined the likely timing of the next nova eruption of T CrB using three complementary constraints: historical recurrence, empirical orbital phase, and recent accretion-state evolution.
Our main conclusions are as follows.

\begin{enumerate}
\item
Using Julian years, the three effective historical intervals have $\bar{T}=79.86$~yr and $s_T=1.53$~yr. Under a descriptive Gaussian recurrence model and conditioning on no eruption by 2026 July 11 (MJD~61232), the probabilities are 30.2\% through the end of 2026 and 56.9\% over the following 12 months. The instantaneous hazard is $0.682~{\rm yr}^{-1}$, but leave-one-out sensitivity ranges are broad; these are empirical indicators, not physical prediction probabilities.

\item
The four historical eruption phases do not define a single preferred ignition phase.
They instead form two loose empirical phase pairs near $\phi\simeq0.44$ and $\phi\simeq0.62$.
After 2026 July 11, the first remaining phase-pair window is 2026 August 5--8.
The future dates associated with the two pairs are monitoring windows, not deterministic predictions, and their dominant uncertainties are systematic.

\item
The 1946 pre-eruption dip is difficult to explain as either a pure accretion-rate decline or a standard foreground-dust event.
If the historical $V$-band depth is physical, the dip likely required both accretion restructuring and source-dependent obscuration.
This makes the 2023--2024 fading episode an imperfect analogue of the pre-1946 dip.

\item
The renewed 2026 decline may test the historical-template interpretation. If it is the true pre-eruption dip, an eruption around 2026 December remains plausible, but the uncertainty is at least several months. This remains a monitoring hypothesis, not a deterministic prediction.

\item
An accretion-deficit estimate suggests that the shorter and fainter 2014--2023 high state could naturally delay the eruption beyond 2026. Figure~\ref{fig:lc} shows that the post-dip phase after 2024 May is fainter on average than the 2014--2023 high state but brighter than the pre-2014 quiescent level, suggesting an intermediate accretion state.
If the mean brightness from 2024 May to MJD~61232, together with the future mean brightness, remains lower than during the 2014--2023 high state, then the continued-high-state compensation time gives an earliest-time lower limit.
Including the small WD-mass correction, this lower limit is approximately 2024 May plus 5.02~yr, i.e., 2029 May.
In this case, the next nova eruption of T CrB would be more likely to occur after 2029 May.

\item
The scenarios are not mutually exclusive predictions. The eventual eruption time will discriminate among their assumptions: a 2026 eruption would favour the dip analogue or a smaller-than-estimated mass deficit; an eruption near or after 2029 under continued sub-high-state brightness would support the accretion-deficit interpretation; and coincidence with a phase window alone would remain insufficient to establish phase locking. The eruption epoch will therefore preserve, rather than remove, the diagnostic value of these tests.
\end{enumerate}

\section*{Data Availability}
The data analyzed in this article are all available in the AAVSO International Database at the following URL: \url{https://www.aavso.org}.

\section*{Acknowledgements}
We extend our sincere gratitude to the anonymous referee for her or his insightful and constructive comments, which helped us to improve the scientific content of this article. We extend our gratitude to observers worldwide for their valuable contributions to the AAVSO database. This work is supported by the High-level Talents Research Start-up Fund Project of Liupanshui Normal University (Grant No. LPSSYKYJJ202208), the Science and Technology Foundation of Guizhou Province (Grant Nos. QKHJC-ZK[2023]442 and QKHJCMS[2026]752), the Discipline-Team of Liupanshui Normal University (Grant No. LPSSY2023XKTD11), the Liupanshui Science and Technology Development Project (Grant Nos. 52020-2024-PT-05, 52020-2025-0-2-08, and 52020-2025-0-2-13), the National Natural Science Foundation of China (Grant Nos. 12303054, 12393853, and 12003020), the Yunnan Fundamental Research Projects (Grant Nos. 202401AU070063 and 202501AS070078), the National Key Research and Development Program of China (Grant No. 2024YFA1611603), the International Centre of Supernovae, Yunnan Key Laboratory (No. 202302AN360001), the Guizhou Provincial Basic Research Program (No. QKHJC[2024]youth158), the High-Level Talent Recruitment Project of Qiannan Normal University for Nationalities (Grant No. qnsyrc202308), and the Shaanxi Fundamental Science Research Project for Mathematics and Physics (Grant No. 23JSY015). R.-Z. Su acknowledges the support from the Shanghai Super Postdoctoral Incentive Program (No. 2025322), the China Postdoctoral Science Foundation (Grant No. 2025M783232), and the National SKA Program of China (No. 2025SKA0130100). Y.-Z. Cai acknowledges financial support from the SOXS project (PI S. Campana).




\bibliographystyle{mnras}
\bibliography{TCrB} 

@ARTICLE{2023A&A...680L..18Z,
       author = {{Zamanov}, R. and {Boeva}, S. and {Latev}, G.~Y. and {Semkov}, E. and {Minev}, M. and {Kostov}, A. and {Bode}, M.~F. and {Marchev}, V. and {Marchev}, D.},
        title = "{Accretion in the recurrent nova T CrB: Linking the superactive state to the predicted outburst}",
      journal = {\aap},
         year = 2023,
        month = dec,
       volume = {680},
          eid = {L18},
        pages = {L18},
          doi = {10.1051/0004-6361/202348372},
archivePrefix = {arXiv},
       eprint = {2312.04342},
 primaryClass = {astro-ph.SR},
       adsurl = {https://ui.adsabs.harvard.edu/abs/2023A&A...680L..18Z}
}

@ARTICLE{1999A&AS..137..473M,
       author = {{M{\"u}rset}, U. and {Schmid}, H.~M.},
        title = "{Spectral classification of the cool giants in symbiotic systems}",
      journal = {\aaps},
         year = 1999,
        month = jun,
       volume = {137},
        pages = {473-493},
          doi = {10.1051/aas:1999105},
       adsurl = {https://ui.adsabs.harvard.edu/abs/1999A&AS..137..473M}
}

@ARTICLE{1998MNRAS.296...77B,
       author = {{Belczynski}, K. and {Mikolajewska}, J.},
        title = "{New binary parameters for the symbiotic recurrent nova T Coronae Borealis}",
      journal = {\mnras},
         year = 1998,
        month = may,
       volume = {296},
       number = {1},
        pages = {77-84},
          doi = {10.1046/j.1365-8711.1998.01301.x},
archivePrefix = {arXiv},
       eprint = {astro-ph/9711151},
 primaryClass = {astro-ph},
       adsurl = {https://ui.adsabs.harvard.edu/abs/1998MNRAS.296...77B}
}

@ARTICLE{2004A&A...415..609S,
       author = {{Stanishev}, V. and {Zamanov}, R. and {Tomov}, N. and {Marziani}, P.},
        title = "{H{\ensuremath{\alpha}} variability of the recurrent nova T Coronae Borealis}",
      journal = {\aap},
         year = 2004,
        month = feb,
       volume = {415},
        pages = {609-616},
          doi = {10.1051/0004-6361:20034623},
archivePrefix = {arXiv},
       eprint = {astro-ph/0311309},
 primaryClass = {astro-ph},
       adsurl = {https://ui.adsabs.harvard.edu/abs/2004A&A...415..609S}
}

@ARTICLE{2023MNRAS.524.3146S,
       author = {{Schaefer}, Bradley E.},
        title = "{The B \& V light curves for recurrent nova T CrB from 1842-2022, the unique pre- and post-eruption high-states, the complex period changes, and the upcoming eruption in 2025.5 {\ensuremath{\pm}} 1.3}",
      journal = {\mnras},
         year = 2023,
        month = sep,
       volume = {524},
       number = {2},
        pages = {3146-3165},
          doi = {10.1093/mnras/stad735},
archivePrefix = {arXiv},
       eprint = {2303.04933},
 primaryClass = {astro-ph.SR},
       adsurl = {https://ui.adsabs.harvard.edu/abs/2023MNRAS.524.3146S}
}

@ARTICLE{1949ApJ...109...81S,
       author = {{Sanford}, Roscoe F.},
        title = "{High-Dispersion Spectrograms of T Coronae Borealis.}",
      journal = {\apj},
         year = 1949,
        month = jan,
       volume = {109},
        pages = {81},
          doi = {10.1086/145106},
       adsurl = {https://ui.adsabs.harvard.edu/abs/1949ApJ...109...81S}
}

@ARTICLE{2023JHA....54..436S,
       author = {{Schaefer}, Bradley E.},
        title = "{The recurrent nova T CrB had prior eruptions observed near December 1787 and October 1217 AD}",
      journal = {Journal for the History of Astronomy},
         year = 2023,
        month = nov,
       volume = {54},
       number = {4},
        pages = {436-455},
          doi = {10.1177/00218286231200492},
archivePrefix = {arXiv},
       eprint = {2308.13668},
 primaryClass = {astro-ph.SR},
       adsurl = {https://ui.adsabs.harvard.edu/abs/2023JHA....54..436S}
}

@ARTICLE{2020ApJ...902L..14L,
       author = {{Luna}, Gerardo J.~M. and {Sokoloski}, J.~L. and {Mukai}, Koji and {M. Kuin}, N. Paul},
        title = "{Increasing Activity in T CrB Suggests Nova Eruption Is Impending}",
      journal = {\apjl},
         year = 2020,
        month = oct,
       volume = {902},
       number = {1},
          eid = {L14},
        pages = {L14},
          doi = {10.3847/2041-8213/abbb2c},
archivePrefix = {arXiv},
       eprint = {2009.11902},
 primaryClass = {astro-ph.SR},
       adsurl = {https://ui.adsabs.harvard.edu/abs/2020ApJ...902L..14L}
}

@ARTICLE{2023ATel16107....1S,
       author = {{Schaefer}, Bradley E. and {Kloppenborg}, Brian and {Waagen}, Elizabeth O. and {Observers}, The Aavso},
        title = "{Recurrent nova T CrB has just started its Pre-eruption Dip in March/April 2023, so the eruption should occur around 2024.4 +- 0.3}",
      journal = {The Astronomer's Telegram},
         year = 2023,
        month = jun,
       volume = {16107},
        pages = {1},
       adsurl = {https://ui.adsabs.harvard.edu/abs/2023ATel16107....1S}
}

@ARTICLE{2022MNRAS.517.6150S,
       author = {{Schaefer}, Bradley E.},
        title = "{Comprehensive catalogue of the overall best distances and properties of 402 galactic novae}",
      journal = {\mnras},
         year = 2022,
        month = dec,
       volume = {517},
       number = {4},
        pages = {6150-6169},
          doi = {10.1093/mnras/stac2900},
archivePrefix = {arXiv},
       eprint = {2210.03181},
 primaryClass = {astro-ph.SR},
       adsurl = {https://ui.adsabs.harvard.edu/abs/2022MNRAS.517.6150S}
}

@ARTICLE{2016NewA...47....7M,
       author = {{Munari}, Ulisse and {Dallaporta}, Sergio and {Cherini}, Giulio},
        title = "{The 2015 super-active state of recurrent nova T CrB and the long term evolution after the 1946 outburst}",
      journal = {\na},
         year = 2016,
        month = aug,
       volume = {47},
        pages = {7-15},
          doi = {10.1016/j.newast.2016.01.002},
archivePrefix = {arXiv},
       eprint = {1602.07470},
 primaryClass = {astro-ph.SR},
       adsurl = {https://ui.adsabs.harvard.edu/abs/2016NewA...47....7M}
}

@ARTICLE{2016MNRAS.462.2695I,
       author = {{I{\l}kiewicz}, Krystian and {Miko{\l}ajewska}, Joanna and {Stoyanov}, Kiril and {Manousakis}, Antonios and {Miszalski}, Brent},
        title = "{Active phases and flickering of a symbiotic recurrent nova T CrB}",
      journal = {\mnras},
         year = 2016,
        month = nov,
       volume = {462},
       number = {3},
        pages = {2695-2705},
          doi = {10.1093/mnras/stw1837},
archivePrefix = {arXiv},
       eprint = {1607.06804},
 primaryClass = {astro-ph.SR},
       adsurl = {https://ui.adsabs.harvard.edu/abs/2016MNRAS.462.2695I}
}

@ARTICLE{2023ApJ...953L...7I,
       author = {{I{\l}kiewicz}, Krystian and {Miko{\l}ajewska}, Joanna and {Stoyanov}, Kiril A.},
        title = "{Symbiotic Star T CrB as an Extreme SU UMa-type Dwarf Nova}",
      journal = {\apjl},
         year = 2023,
        month = aug,
       volume = {953},
       number = {1},
          eid = {L7},
        pages = {L7},
          doi = {10.3847/2041-8213/ace9dc},
archivePrefix = {arXiv},
       eprint = {2307.13838},
 primaryClass = {astro-ph.SR},
       adsurl = {https://ui.adsabs.harvard.edu/abs/2023ApJ...953L...7I}
}

@ARTICLE{2023ATel16114....1K,
       author = {{Kuin}, N. Paul and {Luna}, Gerardo Juan Manuel and {Page}, Kim and {Mukai}, Koji and {Sokoloski}, Jennifer L. and {Osborne}, Julian P. and {Schaefer}, Bradley E.},
        title = "{Swift observations of the changes in the brightness of the recurrent nova T CrB}",
      journal = {The Astronomer's Telegram},
         year = 2023,
        month = jul,
       volume = {16114},
        pages = {1},
       adsurl = {https://ui.adsabs.harvard.edu/abs/2023ATel16114....1K}
}

@ARTICLE{2025MNRAS.541L..14M,
       author = {{Merc}, Jaroslav and {Wyrzykowski}, {\L}ukasz and {Beck}, Paul G. and {Miko{\l}ajczyk}, Przemys{\l}aw J. and {Kotysz}, Krzysztof and {Zieli{\'n}ski}, Pawe{\l} and {Zola}, Staszek and {Kurowski}, Sebastian and {Og{\l}oza}, Waldemar and {Drozdz}, Marek and {Galdies}, Charles and {Hambsch}, Franz-Josef and {Brincat}, Stephen M. and {Joachimczyk}, Barbara and {Bronikowski}, Mateusz and {Japelj}, Jure and {Mihelcic}, Matej and {Carrasco}, Josep Manel and {Burgaz}, Umut and {Gurgul}, Agnieszka and {B{\k{a}}kowska}, Karolina and {Hofbauer}, Piotr and {Szyszka}, Krzysztof and {Golonka}, Jan and {Qvam}, Jan K{\r{a}}re Trandem and {Zdanavi{\v{c}}ius}, Justas and {Pak{\v{s}}tien{\.{e}}}, Erika and {Maskoli{\={u}}nas}, Marius and {{\v{C}}epas}, Vytautas and {Pylypenko}, Uliana and {Mo{\'z}dzierski}, Dawid and {Dubois}, Franky and {Vanaverbeke}, Siegfried and {Olszewska}, Justyna M. and {Liakos}, Alexios and {Stojanovi{\'c}}, Milan and {Damljanovi{\'c}}, Goran and {Popowicz}, Adam and {Marzec}, Mateusz and {Badura}, Magdalena and {Gil}, Bartosz and {Pucek}, Alicja and {Kowalska}, Aleksandra and {Szklarz}, Mateusz and {Kvernadze}, Teimuraz and {Reguitti}, Andrea and {Awiphan}, Supachai and {Dennefeld}, Michel and {Gazeas}, Kosmas},
        title = "{Is the symbiotic recurrent nova T CrB late? Recent photometric evolution and comparison with past pre-outburst behaviour}",
      journal = {\mnras},
         year = 2025,
        month = jul,
       volume = {541},
       number = {1},
        pages = {L14-L21},
          doi = {10.1093/mnrasl/slaf047},
archivePrefix = {arXiv},
       eprint = {2504.20592},
 primaryClass = {astro-ph.SR},
       adsurl = {https://ui.adsabs.harvard.edu/abs/2025MNRAS.541L..14M}
}

@ARTICLE{2025A&A...694A..85P,
       author = {{Planquart}, L. and {Jorissen}, A. and {Van Winckel}, H.},
        title = "{Resolving the mass transfer in the symbiotic recurrent nova T Coronae Borealis}",
      journal = {\aap},
         year = 2025,
        month = feb,
       volume = {694},
          eid = {A85},
        pages = {A85},
          doi = {10.1051/0004-6361/202452833},
archivePrefix = {arXiv},
       eprint = {2501.02984},
 primaryClass = {astro-ph.SR},
       adsurl = {https://ui.adsabs.harvard.edu/abs/2025A&A...694A..85P}
}

@ARTICLE{2018A&A...619A..61L,
       author = {{Luna}, G.~J.~M. and {Mukai}, K. and {Sokoloski}, J.~L. and {Nelson}, T. and {Kuin}, P. and {Segreto}, A. and {Cusumano}, G. and {Jaque Arancibia}, M. and {Nu{\~n}ez}, N.~E.},
        title = "{Dramatic change in the boundary layer in the symbiotic recurrent nova T Coronae Borealis}",
      journal = {\aap},
         year = 2018,
        month = nov,
       volume = {619},
          eid = {A61},
        pages = {A61},
          doi = {10.1051/0004-6361/201833747},
archivePrefix = {arXiv},
       eprint = {1807.01304},
 primaryClass = {astro-ph.HE},
       adsurl = {https://ui.adsabs.harvard.edu/abs/2018A&A...619A..61L}
}

@ARTICLE{2021AJ....161..147B,
       author = {{Bailer-Jones}, C.~A.~L. and {Rybizki}, J. and {Fouesneau}, M. and {Demleitner}, M. and {Andrae}, R.},
        title = "{Estimating Distances from Parallaxes. V. Geometric and Photogeometric Distances to 1.47 Billion Stars in Gaia Early Data Release 3}",
      journal = {\aj},
         year = 2021,
        month = mar,
       volume = {161},
       number = {3},
          eid = {147},
        pages = {147},
          doi = {10.3847/1538-3881/abd806},
archivePrefix = {arXiv},
       eprint = {2012.05220},
 primaryClass = {astro-ph.SR},
       adsurl = {https://ui.adsabs.harvard.edu/abs/2021AJ....161..147B}
}

@ARTICLE{2025ApJ...983...76H,
       author = {{Hinkle}, Kenneth H. and {Nagarajan}, Pranav and {Fekel}, Francis C. and {Miko{\l}ajewska}, Joanna and {Straniero}, Oscar and {Muterspaugh}, Matthew W.},
        title = "{Binary Parameters for the Recurrent Nova T Coronae Borealis}",
      journal = {\apj},
         year = 2025,
        month = apr,
       volume = {983},
       number = {1},
          eid = {76},
        pages = {76},
          doi = {10.3847/1538-4357/adbe63},
archivePrefix = {arXiv},
       eprint = {2502.20664},
 primaryClass = {astro-ph.SR},
       adsurl = {https://ui.adsabs.harvard.edu/abs/2025ApJ...983...76H}
}

@ARTICLE{1975JBAA...85..217B,
       author = {{Bailey}, J.},
        title = "{Periodic fluctuations in the recurrent nova T CrB.}",
      journal = {Journal of the British Astronomical Association},
         year = 1975,
        month = mar,
       volume = {85},
        pages = {217-223},
       adsurl = {https://ui.adsabs.harvard.edu/abs/1975JBAA...85..217B}
}

@ARTICLE{2023BAAVC.196....8T,
       author = {{Toone}, John},
        title = "{T Coronae Borealis {\textendash} The Pre-Eruption Dip}",
      journal = {British Astronomical Association Variable Star Section Circular},
         year = 2023,
        month = jun,
       volume = {196},
        pages = {8-9},
       adsurl = {https://ui.adsabs.harvard.edu/abs/2023BAAVC.196....8T}
}

@ARTICLE{2000AJ....119.1375F,
       author = {{Fekel}, Francis C. and {Joyce}, Richard R. and {Hinkle}, Kenneth H. and {Skrutskie}, Michael F.},
        title = "{Infrared Spectroscopy of Symbiotic Stars. I. Orbits for Well-Known S-Type Systems}",
      journal = {\aj},
         year = 2000,
        month = mar,
       volume = {119},
       number = {3},
        pages = {1375-1388},
          doi = {10.1086/301260},
       adsurl = {https://ui.adsabs.harvard.edu/abs/2000AJ....119.1375F}
}

@ARTICLE{2023AstL...49..501M,
       author = {{Maslennikova}, N.~A. and {Tatarnikov}, A.~M. and {Tatarnikova}, A.~A. and {Dodin}, A.~V. and {Shenavrin}, V.~I. and {Burlak}, M.~A. and {Zheltoukhov}, S.~G. and {Strakhov}, I.~A.},
        title = "{Recurrent Symbiotic Nova T Coronae Borealis before Outburst}",
      journal = {Astronomy Letters},
         year = 2023,
        month = oct,
       volume = {49},
       number = {9},
        pages = {501-515},
          doi = {10.1134/S1063773723090037},
archivePrefix = {arXiv},
       eprint = {2308.10011},
 primaryClass = {astro-ph.SR},
       adsurl = {https://ui.adsabs.harvard.edu/abs/2023AstL...49..501M}
}

@ARTICLE{2025ApJ...989...78S,
       author = {{Schlindwein}, Wagner and {Baptista}, Raymundo and {Luna}, Gerardo Juan Manuel},
        title = "{Modeling the High-brightness State of the Recurrent Nova T CrB as an Enhanced Mass-transfer Event}",
      journal = {\apj},
         year = 2025,
        month = aug,
       volume = {989},
       number = {1},
          eid = {78},
        pages = {78},
          doi = {10.3847/1538-4357/ade98c},
archivePrefix = {arXiv},
       eprint = {2506.05098},
 primaryClass = {astro-ph.SR},
       adsurl = {https://ui.adsabs.harvard.edu/abs/2025ApJ...989...78S}
}

@ARTICLE{2025A&A...701A.176M,
       author = {{Munari}, U. and {Walter}, F. and {Masetti}, N. and {Valisa}, P. and {Dallaporta}, S. and {Bergamini}, A. and {Cherini}, G. and {Frigo}, A. and {Maitan}, A. and {Marino}, C. and {Mazzacurati}, G. and {Moretti}, S. and {Tabacco}, F. and {Tomaselli}, S. and {Vagnozzi}, A. and {Ochner}, P. and {Albanese}, I.},
        title = "{T CrB: Overview of the accretion history, Roche-lobe filling, orbital solution, and radiative modeling}",
      journal = {\aap},
         year = 2025,
        month = sep,
       volume = {701},
          eid = {A176},
        pages = {A176},
          doi = {10.1051/0004-6361/202555917},
       adsurl = {https://ui.adsabs.harvard.edu/abs/2025A&A...701A.176M}
}

@ARTICLE{2024RNAAS...8..272S,
       author = {{Schneider}, Jean},
        title = "{When will the Next T CrB Eruption Occur?}",
      journal = {Research Notes of the American Astronomical Society},
         year = 2024,
        month = oct,
       volume = {8},
       number = {10},
          eid = {272},
        pages = {272},
          doi = {10.3847/2515-5172/ad8bba},
       adsurl = {https://ui.adsabs.harvard.edu/abs/2024RNAAS...8..272S}
}

@ARTICLE{2026A&A...706A..94P,
       author = {{Pei}, Songpeng and {Zhang}, Xiaowan and {Su}, Renzhi and {Cai}, Yongzhi and {Ou}, Ziwei and {Li}, Qiang and {Ren}, Xiaoqin and {Yang}, Taozhi and {Li}, Mingyue},
        title = "{Multiwavelength study of the pre-eruption dip in the recurrent nova T Coronae Borealis preceding imminent nova eruption}",
      journal = {\aap},
         year = 2026,
        month = feb,
       volume = {706},
          eid = {A94},
        pages = {A94},
          doi = {10.1051/0004-6361/202557346},
archivePrefix = {arXiv},
       eprint = {2512.19218},
 primaryClass = {astro-ph.HE},
       adsurl = {https://ui.adsabs.harvard.edu/abs/2026A&A...706A..94P}
}

@ARTICLE{2026A&A...707A.102L,
       author = {{Luna}, G.~J.~M. and {Kuin}, N.~P.~M. and {Mukai}, K. and {Sokoloski}, J.~L. and {Page}, K. and {Osborne}, J.~P.},
        title = "{Evolution of the recent high-accretion state of the recurrent nova T CrB: HST, Swift, NuSTAR, and XMM-Newton observations}",
      journal = {\aap},
         year = 2026,
        month = feb,
       volume = {707},
          eid = {A102},
        pages = {A102},
          doi = {10.1051/0004-6361/202557435},
archivePrefix = {arXiv},
       eprint = {2601.16190},
 primaryClass = {astro-ph.HE},
       adsurl = {https://ui.adsabs.harvard.edu/abs/2026A&A...707A.102L}
}

@ARTICLE{2025ApJ...991..111S,
       author = {{Schaefer}, Bradley E.},
        title = "{Orbital Period Changes in Recurrent Nova T Corona Borealis Prove That It Is Not a Type Ia Supernovae Progenitor}",
      journal = {\apj},
         year = 2025,
        month = sep,
       volume = {991},
       number = {1},
          eid = {111},
        pages = {111},
          doi = {10.3847/1538-4357/adfa16},
archivePrefix = {arXiv},
       eprint = {2510.01587},
 primaryClass = {astro-ph.SR},
       adsurl = {https://ui.adsabs.harvard.edu/abs/2025ApJ...991..111S}
}

@ARTICLE{1982ApJ...257..752F,
       author = {{Fujimoto}, M.~Y.},
        title = "{A theory of hydrogen shell flashes on accreting white dwarfs. I - Their progress and the expansion of the envelope. II - The stable shell burning and the recurrence period of shell flashes}",
      journal = {\apj},
         year = 1982,
        month = jun,
       volume = {257},
        pages = {752-779},
          doi = {10.1086/160029},
       adsurl = {https://ui.adsabs.harvard.edu/abs/1982ApJ...257..752F}
}

@ARTICLE{1982ApJ...257..767F,
       author = {{Fujimoto}, M.~Y.},
        title = "{A Theory of Hydrogen Shell Flashes on Accreting White Dwarfs - Part Two - the Stable Shell Burning and the Recurrence Period of Shell Flashes}",
      journal = {\apj},
         year = 1982,
        month = jun,
       volume = {257},
        pages = {767},
          doi = {10.1086/160030},
       adsurl = {https://ui.adsabs.harvard.edu/abs/1982ApJ...257..767F}
}

@ARTICLE{2004ApJ...600..390T,
       author = {{Townsley}, Dean M. and {Bildsten}, Lars},
        title = "{Theoretical Modeling of the Thermal State of Accreting White Dwarfs Undergoing Classical Nova Cycles}",
      journal = {\apj},
         year = 2004,
        month = jan,
       volume = {600},
       number = {1},
        pages = {390-403},
          doi = {10.1086/379647},
archivePrefix = {arXiv},
       eprint = {astro-ph/0306080},
 primaryClass = {astro-ph},
       adsurl = {https://ui.adsabs.harvard.edu/abs/2004ApJ...600..390T}
}

@ARTICLE{2005ApJ...623..398Y,
       author = {{Yaron}, O. and {Prialnik}, D. and {Shara}, M.~M. and {Kovetz}, A.},
        title = "{An Extended Grid of Nova Models. II. The Parameter Space of Nova Outbursts}",
      journal = {\apj},
         year = 2005,
        month = apr,
       volume = {623},
       number = {1},
        pages = {398-410},
          doi = {10.1086/428435},
archivePrefix = {arXiv},
       eprint = {astro-ph/0503143},
 primaryClass = {astro-ph},
       adsurl = {https://ui.adsabs.harvard.edu/abs/2005ApJ...623..398Y}
}

@ARTICLE{2018ApJ...860..110S,
       author = {{Shara}, Michael M. and {Prialnik}, Dina and {Hillman}, Yael and {Kovetz}, Attay},
        title = "{The Masses and Accretion Rates of White Dwarfs in Classical and Recurrent Novae}",
      journal = {\apj},
         year = 2018,
        month = jun,
       volume = {860},
       number = {2},
          eid = {110},
        pages = {110},
          doi = {10.3847/1538-4357/aabfbd},
archivePrefix = {arXiv},
       eprint = {1804.06880},
 primaryClass = {astro-ph.SR},
       adsurl = {https://ui.adsabs.harvard.edu/abs/2018ApJ...860..110S}
}

@ARTICLE{2019ApJS..242...18H,
       author = {{Hachisu}, Izumi and {Kato}, Mariko},
        title = "{A Light-curve Analysis of 32 Recent Galactic Novae: Distances and White Dwarf Masses}",
      journal = {\apjs},
         year = 2019,
        month = jun,
       volume = {242},
       number = {2},
          eid = {18},
        pages = {18},
          doi = {10.3847/1538-4365/ab1b43},
archivePrefix = {arXiv},
       eprint = {1905.10655},
 primaryClass = {astro-ph.SR},
       adsurl = {https://ui.adsabs.harvard.edu/abs/2019ApJS..242...18H}
}

@ARTICLE{2009ApJ...692..324S,
       author = {{Shen}, Ken J. and {Bildsten}, Lars},
        title = "{The Effect of Composition on Nova Ignitions}",
      journal = {\apj},
         year = 2009,
        month = feb,
       volume = {692},
       number = {1},
        pages = {324-334},
          doi = {10.1088/0004-637X/692/1/324},
archivePrefix = {arXiv},
       eprint = {0805.2160},
 primaryClass = {astro-ph},
       adsurl = {https://ui.adsabs.harvard.edu/abs/2009ApJ...692..324S}
}

@ARTICLE{2026arXiv260520991P,
       author = {{Pei}, Songpeng and {Zhang}, Xiaowan and {Su}, Renzhi and {Cai}, Yongzhi and {Ou}, Ziwei and {Li}, Qiang and {Ren}, Xiaoqin and {Liu}, Yu and {Yang}, Taozhi},
        title = "{Lack of Significant Orbital-Phase Locking in the Active Phases of the Recurrent Nova T CrB}",
      journal = {arXiv e-prints},
         year = 2026,
        month = may,
          eid = {arXiv:2605.20991},
        pages = {arXiv:2605.20991},
archivePrefix = {arXiv},
       eprint = {2605.20991},
 primaryClass = {astro-ph.HE},
       adsurl = {https://ui.adsabs.harvard.edu/abs/2026arXiv260520991P}
}

@ARTICLE{1989ApJ...345..245C,
       author = {{Cardelli}, Jason A. and {Clayton}, Geoffrey C. and {Mathis}, John S.},
        title = "{The Relationship between Infrared, Optical, and Ultraviolet Extinction}",
      journal = {\apj},
         year = 1989,
        month = oct,
       volume = {345},
        pages = {245},
          doi = {10.1086/167900},
       adsurl = {https://ui.adsabs.harvard.edu/abs/1989ApJ...345..245C}
}

@ARTICLE{1999PASP..111...63F,
       author = {{Fitzpatrick}, Edward L.},
        title = "{Correcting for the Effects of Interstellar Extinction}",
      journal = {\pasp},
         year = 1999,
        month = jan,
       volume = {111},
       number = {755},
        pages = {63-75},
          doi = {10.1086/316293},
archivePrefix = {arXiv},
       eprint = {astro-ph/9809387},
 primaryClass = {astro-ph},
       adsurl = {https://ui.adsabs.harvard.edu/abs/1999PASP..111...63F}
}

@ARTICLE{1972ApJ...175..417N,
       author = {{Nauenberg}, Michael},
        title = "{Analytic Approximations to the Mass-Radius Relation and Energy of Zero-Temperature Stars}",
      journal = {\apj},
         year = 1972,
        month = jul,
       volume = {175},
        pages = {417},
          doi = {10.1086/151568},
       adsurl = {https://ui.adsabs.harvard.edu/abs/1972ApJ...175..417N}
}




\bsp	
\label{lastpage}
\end{document}